\documentclass[pra,twocolumn,showpacs,amsfonts,amsmath,floatfix]{revtex4}
\usepackage{epsfig,amsmath,graphicx}

\usepackage{graphicx}
\usepackage{dcolumn}
\usepackage{bm}
\usepackage{hyperref}

\begin{document}


\title{Frequency stabilization of the non resonant wave of a continuous-wave singly resonant optical parametric oscillator}

\author{Aliou Ly$^{1}$}
\author{Benjamin Szymanski$^{2}$ }
\author{Fabien Bretenaker$^{1}$}
\email{Fabien.Bretenaker@u-psud.fr}
 \affiliation {$^1$Laboratoire Aim\'e Cotton, CNRS-ENS Cachan-Universit\'e Paris Sud 11, 91405 Orsay Cedex, France\\$^2$Blue Industry \& Science, 208 bis rue la Fayette, 75010 Paris, France}

\date{\today}

\begin{abstract}
We present an experimental technique allowing to stabilize the frequency of the non resonant wave in a singly resonant optical parametric oscillator (SRO) down to the kHz level, much below the pump frequency noise level. By comparing the frequency of the non resonant wave with a reference cavity, the pump frequency noise is imposed to the frequency of the resonant wave, and is thus subtracted from the frequency of the non resonant wave. This permits the non resonant wave obtained from such a SRO to be simultaneously powerful and frequency stable, which is usually impossible to obtain when the resonant wave frequency is stabilized.
\end{abstract}

\maketitle
\section{Introduction}
\label{intro}
Continuous -- wave (CW) optical parametric oscillators  (OPOs) are versatile sources of coherent light. They permit to reach single -- frequency operation on a very broad wavelength tuning range, and multi-Watt output power, which makes them well suited for applications to atomic physics \cite{Zaske10,Gross10} and high resolution molecular spectroscopy \cite{Kovalchuk01}, among other applications \cite{Breuning11}. Concerning their output power, singly resonant OPOs (SROs) have long been sought for their high efficiency, as predicted by earliest theoretical studies \cite{Kreuzer69,Harris69}. Indeed, in the plane wave approximation, a $100\,\%$ conversion efficiency is predicted for pumping at $(\pi/2)^2$ times above the oscillation threshold. This theoretical limit was almost reached with a CW SRO based on a periodically poled lithium niobate (PPLN) crystal for which the authors reported $93\,\%$ of pump depletion \cite{Bosenberg96b}. In most cases, the optimization of the output power depends on the choices of the crystal length and mirror reflectivities for a given available pump power. A recent example shows that a low threshold is not necessarily the best way to obtain a high output power, even for the non resonant beam (idler) \cite{Sowade09}. 

Besides, SROs are also extremely attractive for their frequency noise properties. In particular, it has been shown that the frequency noise of the pump can be dumped to the wave which is not resonant inside the cavity \cite{Mhibik10}. By locking the frequency of the resonant wave at resonance with a high finesse resonator, this has allowed to stabilize this frequency down to the kHz level \cite{Mhibik11,Andrieux11}, i. e., well below the linewidth of the pump laser. The problem then is that the output power obtained from the resonant wavelength is usually quite small, i.~e. well below 1 W for pump powers of several Watts. Indeed, the converted pump power is mostly extracted by the non resonant wave which carries the pump noise.  This problem has received quite some attention. For example, extracting some power from the wave resonant in the cavity can be performed by using an optimized output coupler such as a dielectric coating mirror \cite{Samanta08,Gross10} or a volume Bragg grating \cite{Vainio10}. Another method  consists in inserting a  plate inside the cavity at an incidence angle close to the Brewster angle in order to ensure a small but adjustable amount of output coupling \cite{Paboeuf12}. 

However, in spite of the partial success of these demonstrations, it would be extremely desirable to be able to stabilize the frequency of the non resonating wave. This would allow us to benefit from the two main advantages of SROs -- efficient conversion and extremely small linewidth -- in the same wave. To this aim, we explore here the possibility to add to the resonant wave the amount of frequency noise that would allow it to mimic the frequency fluctuations of the pump. Then, thanks to energy conservation during the parametric process, the frequency of the non resonant wave should be immune of noise. We thus describe in the following how we implemented such a stabilization technique and the obtained results. 

\section{Principle of the frequency stabilization scheme}\label{principle}
\begin{figure}
\resizebox{0.45\textwidth}{!}{%
  \includegraphics{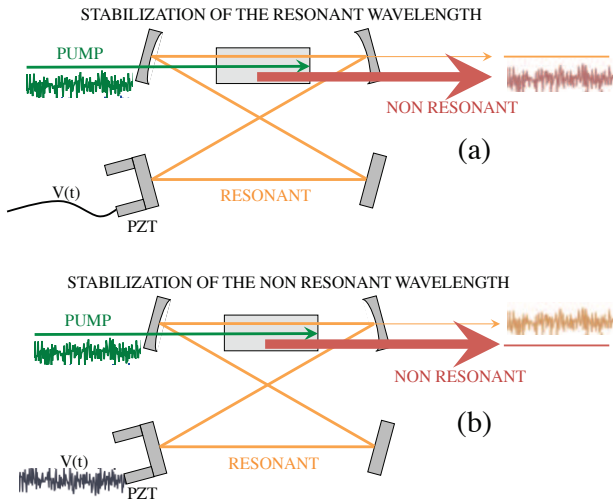}}
  \vspace*{0cm} 
\caption{Principle of the stabilization of the frequency of a SRO. (a) The frequency of the resonant wave is stabilized and the pump frequency noise is transferred to the non resonant one. (b) The frequency of the non resonant wave is stabilized by imposing the pump frequency noise on the resonant one. }
\label{Fig01} 
\end{figure}

The principle of the frequency stabilization scheme tested here is schematized in Fig.\ \ref{Fig01}. In such a parametric oscillator, the energy conservation between the three beams reads
\begin{equation}
\omega_{\mathrm{p}}(t)=\omega_{\mathrm{r}}(t)+\omega_{\mathrm{nr}}(t)\ ,\label{eq01}
\end{equation}
where $\omega_{\mathrm{p}}$, $\omega_{\mathrm{r}}$, and $\omega_{\mathrm{nr}}$ are the instantaneous frequencies of the pump, the wave resonant in the cavity (which can be the signal or the idler), and the non resonant wave (idler or signal, respectively). We suppose that these frequencies vary with time, i. e., contain noises, and that the instantaneous frequencies of these noises are slow enough, compared to the OPO oscillation building time, for eq.~(\ref{eq01}) to be instantaneously satisfied. Now, the frequency of the resonant wave is constrained to be resonant in the cavity of optical length $L$, leading to:
\begin{equation}
\omega_{\mathrm{r}}(t)=2\pi p \frac{c}{L(t)}\ ,\label{eq02}
\end{equation}
where $p$ is an integer.

Let us then consider the configuration of Fig.\ \ref{Fig01}(a), which corresponds to the stabilization scheme developed in Refs.\ \cite{Mhibik10,Mhibik11}. In order to stabilize the frequency of the resonant wave, the cavity length $L$ in eq.\ (\ref{eq02}) must be kept as stable as possible. According to eq.~(\ref{eq01}), the frequency fluctuations $\delta\omega_{\mathrm{p}}(t)$ of the pump are transferred to the non resonant wave only:
\begin{eqnarray}
\delta\omega_{\mathrm{nr}}(t)&=&\delta\omega_{\mathrm{p}}(t)\ ,\label{eq03}\\
\delta\omega_{\mathrm{r}}(t)&=&0\ .\label{eq04}
\end{eqnarray}
As stated in the introduction, this scheme has been very successful in stabilizing the frequency of the resonant signal. However, the wave that carries the largest output power, i. e. the non resonant one, receives all the pump noise. Conversely, if now we want to stabilize the frequency of the non resonant wave, as schematized in Fig.~\ref{Fig01}(b), eqs.~(\ref{eq03}) and (\ref{eq04}) become:
\begin{eqnarray}
\delta\omega_{\mathrm{nr}}(t)&=&0\ ,\label{eq05}\\
\delta\omega_{\mathrm{r}}(t)&=&\delta\omega_{\mathrm{p}}(t)\ .\label{eq06}
\end{eqnarray}
Eq.~(\ref{eq02}) thus shows that some noise $\delta L(t)$ must be introduced in the cavity length for the frequency noise of the resonant wave to mimic the pump one:
\begin{equation}
\frac{\delta L(t)}{L}=-\frac{\delta\omega_{\mathrm{r}}(t)}{\omega_{\mathrm{r}}}=-\frac{\delta\omega_{\mathrm{p}}(t)}{\omega_{\mathrm{r}}}\ .\label{eq07}
\end{equation}
This is the scheme which is implemented in the experiment described below.

\section{Experimental setup}
\label{setup}

\begin{figure}
\resizebox{0.5\textwidth}{!}{%
  \includegraphics{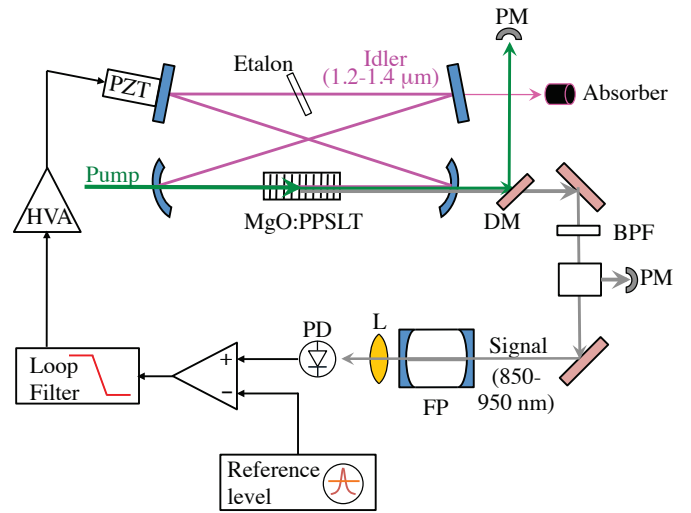}}
  \vspace*{0cm} 
\caption{Experimental setup. HVA: high-voltage amplifier. PZT: piezoelectric transducer. PD: Photodiode. L: focusing lens. FP: Fabry-Perot cavity. DM: Dichroic Mirror. BPF: Bandpass Filter. NPBC: Non Polarizing Beamsplitter Cube. PM: Power Meter. }
\label{Fig02} 
\end{figure}

The experimental setup is depicted in Fig.\,\ref{Fig02}. The OPO, which is similar to the one used in Ref.~\cite{Mhibik11}, is pumped at 532~nm by a cw 10~W single-frequency Coherent Verdi  laser and is based on a 30-mm long MgO-doped periodically poled stochiometric lithium tantalate (PPSLT) crystal ($d_{\mathrm{eff}}\simeq11~\mathrm{pm/V}$) manufactured and coated by HC Photonics. This crystal contains a single grating with a period of 7.97~$\mu$m and is anti-reflection coated for the pump, the signal, and the idler. It is designed to lead to quasi-phase matching conditions for an idler wavelength in the 1200-1400~nm range, depending on the temperature. The OPO cavity is a 1.15-m long ring cavity and consists in four mirrors. The two mirrors that sandwich the nonlinear crystal both have a 150~mm radius of curvature. The two other mirrors are planar with one of them acting as an output coupler for the resonant wave. All mirrors are designed to exhibit a reflectivity larger than 99.8~\% between 1.2~$\mu$m and 1.4~$\mu$m and a transmission larger than 95~\% at 532~nm and between 850~nm and 950~nm. This allows  the OPO to be singly resonant, with the resonant wave of Sec. \ref{principle} being the idler. The estimated waist of the idler beam at the middle of the PPSLT crystal is 37~$\mu$m. The pump beam is focused to a 53~$\mu$m waist inside the PPSLT crystal. 

\begin{figure}
\resizebox{0.5\textwidth}{!}{%
  \includegraphics{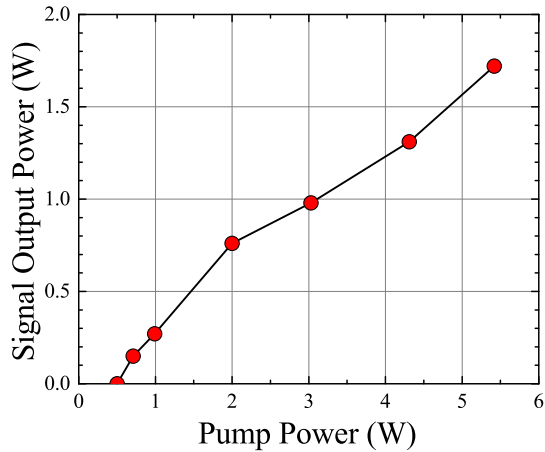}}
  \vspace*{0cm} 
\caption{Measured evolution of the signal output power versus pump power. }
\label{Fig02N1} 
\end{figure}

A 150~$\mu$m thick uncoated Nd:YAG \'etalon  is inserted in the second waist of the cavity to ensure stable single-frequency operation of the OPO. Heating the PPSLT crystal at T=100\,$^{\circ}$~C, we measure, with a spectrometer (AvaSpec 2048-2) not represented here, an idler wavelength of 1213\,nm and a signal wavelength of 947\,nm. The OPO threshold corresponds to an input pump power $P_{\mathrm{p}}^{\mathrm{(in)}}=500\,\mathrm{mW}$. The evolution of the signal output power versus pump power is reproduced in Fig.\,\ref{Fig02N1}. One can clearly see that the extracted pump power is mainly converted to non resonant signal, leading to an output power larger than 1~Watt.

The OPO output beams are collimated with a  300\,mm focal length lens, not represented in Fig.\,\ref{Fig02}. The  dichroic mirror separates the pump from the signal beam.  The pump power $P_{\mathrm{p}}^{\mathrm{(out)}}$ at the output of the OPO is measured to evaluate the pump depletion $\eta= 1-P_{\mathrm{p}}^{\mathrm{(out)}}/P_{\mathrm{p}}^{\mathrm{(in)}}$. Since the dichroic mirror is not perfect, we introduce a bandpass filter centered at 950\,nm, with a bandwith of 10\,nm, on the signal path, to make sure that only  the signal is sent into the Fabry-Perot cavity.

\begin{figure}
\resizebox{0.5\textwidth}{!}{%
  \includegraphics{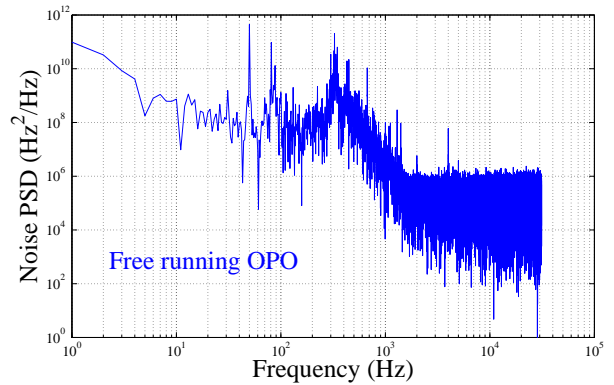}}
  \vspace*{0cm} 
\caption{Single-sided power spectral density of the frequency noise of the non resonant signal obtained from the free running SRO. }
\label{Fig03} 
\end{figure}

In order to optimize the frequency locking of the OPO, we first perform a measurement of the frequency noise spectrum of the non resonant signal beam. To this aim, we use a low finesse Fabry-Perot cavity as a frequency to intensity noise converter \cite{Mhibik10}. This first cavity has a 750~MHz free spectral range and a finesse $F=28$ at the signal wavelength of 947\,nm.
The OPO signal frequency is tuned the side of the transmission peak of this analysis cavity, and the transmitted signal is recorded during 1\,s using a deep memory oscilloscope. A fast Fourier transform algorithm is then used to retrieve the frequency noise spectrum of the non resonant signal, which is reproduced in Fig.\,\ref{Fig03}. This measurement has been obtained for a pump power $P_{\mathrm{p}}^{\mathrm{(in)}}=2.5\,\mathrm{W}$, corresponding to a pump depletion equal to $\eta=80\,\%$ and a signal power of the order of 800\,mW.
This spectrum reproduces fairly well the spectrum of the pump frequency noise \cite{Mhibik10}, showing that the pump noise is transferred to the non resonant signal. Integrating this noise power spectral density over all frequencies, we obtain a RMS frequency noise of 1.3\,MHz. Moreover, we checked that the contributions of the detection noise and the signal intensity noise are negligible. The spectrum of Fig.\,\ref{Fig03} shows that the frequency noise essentially lies at low frequencies (below 1 kHz), thus validating the hypothesis of eq.~(\ref{eq01}). The plateau observed above 1~kHz corresponds to the detection limit, which is pretty high due to the relatively poor linewidth ($\approx27\,\mathrm{MHz}$) of the cavity. 

In order to stabilize the frequency of the non resonant signal wave, we must apply the noise measured in Fig.\,\ref{Fig03} to the cavity length, in order to satisfy eq.\,(\ref{eq07}). To do this with a better signal-to-noise ratio than the one of Fig.\,\ref{Fig03}, we turn to a Fabry-Perot cavity with a better linewidth. This reference cavity has a free spectral range of 1~GHz and a finesse $F=100$ at 947\,nm, leading to a linewidth of the order of 10~MHz. In order to build the error signal that will drive the cavity length variations, the intensity at the output of the cavity is detected, and the corresponding signal is subtracted from a reference voltage. This reference voltage is adjusted in such a way that a zero error signal corresponds to the situation where the signal frequency is tuned half way between the minimum and the maximum transmission of the cavity, where the cavity response is almost linear and exhibits a maximum slope \cite{Vassen90}. The error signal is filtered by a proportional-integral (PI) loop filter, amplified through a high-voltage amplifier, before being applied to a piezoelectric transducer (Piezomechanik model PSt 150/10x10/2) carrying one the cavity planar mirrors (see Fig.\,\ref{Fig02}).

\begin{figure}
\resizebox{0.5\textwidth}{!}{%
  \includegraphics{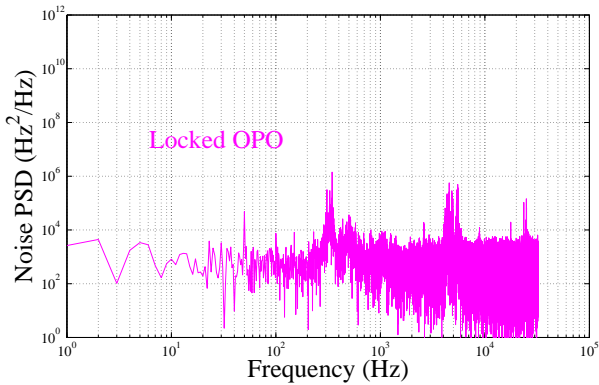}}
  \vspace*{0cm} 
\caption{Single-sided power spectral density of the frequency noise of the non resonant signal obtained from the frequency locked SRO. }
\label{Fig04} 
\end{figure}

To lock the OPO signal frequency, we set the PI corner frequency of the servo controller at 10\,kHz. In these conditions, once the loop is closed, we record the error signal during 1~s and process it in the same way as before in order to retrieve the power spectral density of the frequency noise of the OPO signal. The result is reproduced in Fig.\,\ref{Fig04}.

By comparing Figs.\,\ref{Fig03} and \ref{Fig04}, we see that the low frequency noise is reduced by 8 orders of magnitude when the SRO is locked. This leads to a nearly white frequency noise in the considered frequency domain. The remaining small peaks, at frequencies between 100\,Hz and 1\,kHz, are probably due to electrical perturbations and/or to the fluctuations of the pump frequency. However, there is a remarkable peak around the 4\,kHz frequency which reaches almost $10^6\,\mathrm{Hz}^2/\mathrm{Hz}$. This peak probably comes from the fact that the loop becomes unstable at such frequencies, and could probably be suppressed by a better management of the servo-loop phase. By integrating this noise power spectral density from 1\,Hz to 30\,kHz, we obtain a RMS frequency noise of 7.3\,kHz. This value has to be compared with the value of 1.3\,MHz obtained previously for the pump laser and the spectrum of Fig.\,\ref{Fig03}. The important conclusion of this result is that we were able to dump the frequency noise of the pump into the resonant wave, as expected from eqs.\,(\ref{eq05},\ref{eq06}).

\begin{figure}
\resizebox{0.5\textwidth}{!}{%
  \includegraphics{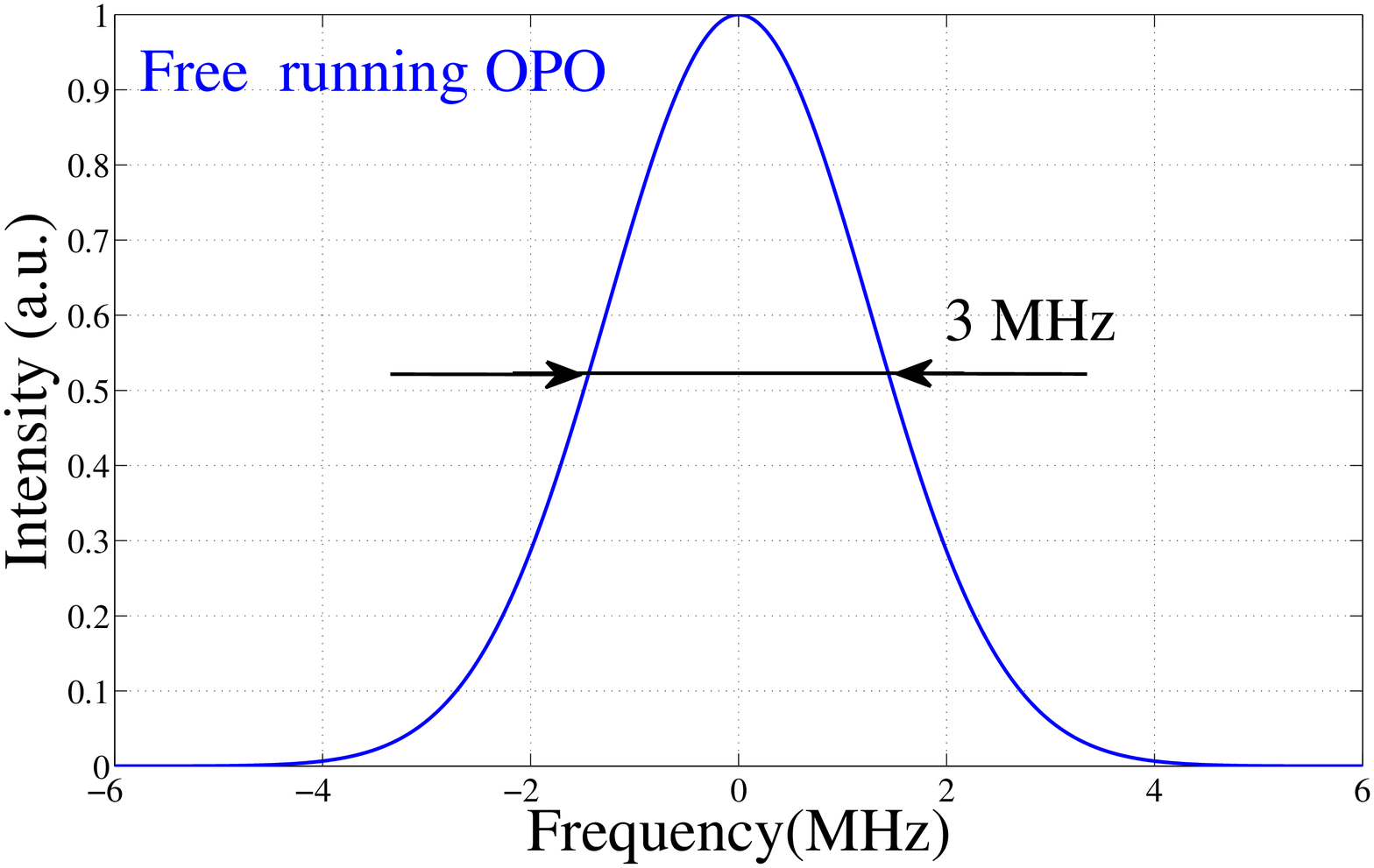}}
  \vspace*{0cm} 
\caption{Lineshape of the non resonant signal of the SRO in the free running mode. }
\label{Fig05} 
\end{figure}

\begin{figure}
\resizebox{0.5\textwidth}{!}{%
  \includegraphics{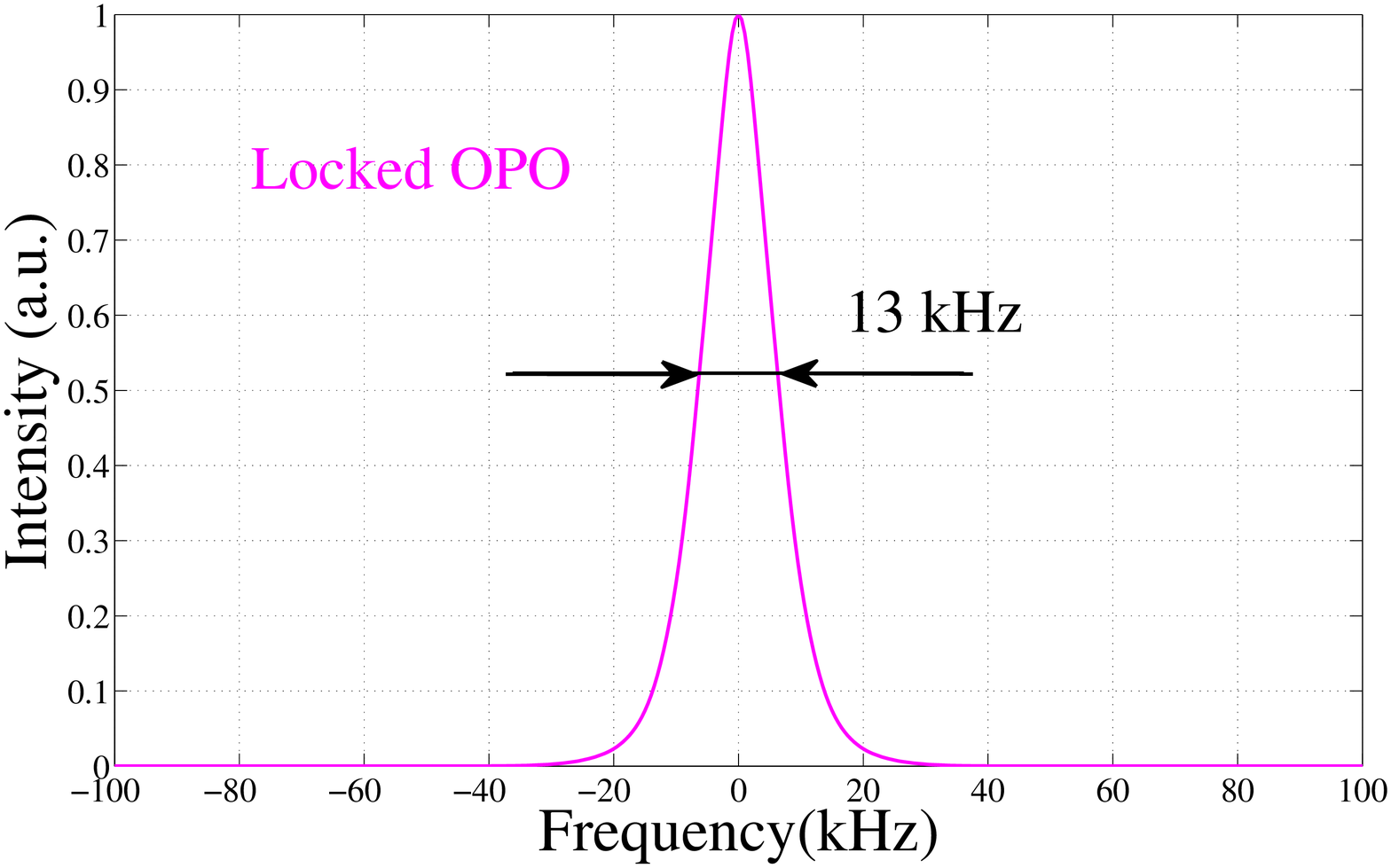}}
  \vspace*{0cm} 
\caption{Lineshape of the non resonant signal of the SRO in the locked mode. }
\label{Fig06} 
\end{figure}

In order to have an idea of the signal linewidth, we have also numerically calculated its spectrum when the OPO is in the free running mode and when it is locked. This is performed by neglecting the OPO intensity noise and by calculating the auto-correlation of the field $E$ of the OPO signal using the following expression \cite{Uehara95}:
\begin{equation}
R_E(\tau)\propto\exp\left[-\int_{-\infty}^{\infty}S_{\nu}(f)\frac{1-\cos (2\pi f \tau)}{f}\mathrm{d}f\right]\ ,\label{eq08}
\end{equation}
where $S_{\nu}(f)$ is the power spectral density of the frequency noise. By injecting the results of Figs.\,\ref{Fig03} and \ref{Fig04} into eq.\,(\ref{eq08}) and performing the Fourier transform of $R_E(\tau)$, we obtain the laser spectra of Figs.\,\ref{Fig05} and \ref{Fig06}, respectively.  

The spectrum of Fig.\ \ref{Fig06} has a Lorentzian lineshape, which is consistent with the fact that the corresponding frequency noise spectra is almost white (see Fig.\ \ref{Fig04}). Moreover, we can see that by adding noise to the resonant wave, we have been able to decrease the linewidth of the non resonant signal by more than three orders of magnitude, down to the kHz domain.

\section{Conclusion}
\label{Conclusion}
In conclusion, we have demonstrated, for the first time to our knowledge, the possibility to lock the frequency of the non resonant wave of a singly resonant OPO down to a frequency noise much lower that the pump frequency noise. This technique opens interesting potentialities in order to obtain both  high power and spectrally pure output, contrary to the use of the resonant wave which can provide only much lower output powers. The present experiment, which aimed at demonstrating the validity of the approach, led to a power of the order of 1 Watt with a relative noise corresponding to a linewidth of the order of 10 kHz. Future steps will include the optimization of the OPO cavity losses in order to obtain several Watts of output power, as evidenced in Ref.\ \cite{My08}, together with the development of a Pound-Drever-Hall stabilization scheme with a higher finesse cavity \cite{Drever83,Salomon88,Mhibik11}  in order to reach the sub-kilohertz linewidth level for the non resonant wave.

\end{document}